\documentclass[prb,aps,floats,twocolumn,showpacs,groupedaddress,floatfix,amsmath]{revtex4-1}
\usepackage[pdftex]{graphicx}
\usepackage{multirow}
\usepackage{dcolumn}
\usepackage{bm}
\def\bc{\begin{center}}
\def\ec{\end{center}}
\def\be{\begin{equation}}
\def\ee{\end{equation}}
\usepackage{longtable}
\usepackage{amsmath}
\usepackage{amssymb}
\usepackage{bm}
\renewcommand{\vec}[1]{\mbox{\boldmath$#1$}}
\begin{document}
\title{Mechanism for current saturation and energy dissipation in graphene transistors}
\author{Ashley M. DaSilva}
\thanks{These authors contributed equally.}
\author{Ke Zou}
\thanks{These authors contributed equally.}
\author{J. K. Jain}
\author{Jun Zhu}
\affiliation{Department of Physics, 104 Davey Laboratory, The Pennsylvania State University, University Park, Pennsylvania 16802}
\begin{abstract}
From a combination of careful and detailed theoretical and experimental studies, we demonstrate that the Boltzmann theory including all scattering mechanisms gives an excellent account, with no adjustable parameters, of high electric field transport in single as well as double-oxide graphene transistors. We further show 
unambiguously that scattering from the substrate and superstrate surface optical (SO) phonons governs the high field transport and heat dissipation over a wide range of experimentally relevant parameters. Models that neglect SO phonons altogether or treat them in a simple phenomenological manner are inadequate. We outline possible strategies for achieving higher current and complete saturation in graphene devices.
\end{abstract}

\maketitle

Electronic devices operating in the radio frequency (RF) regime play a central role in modern communication. Graphene possesses a compelling potential in high-efficiency RF analog devices due to its excellent carrier mobility $\mu >$ 10,000 cm$^2$/Vs at room temperature. Critical to the operation of such devices is the realization and control of carrier velocity saturation at high source-drain electric field in graphene transistors. This subject has attracted numerous experimental and theoretical studies recently~\cite{Meric2008, Bistritzer2009a, Barreiro2009, Freitag2009,Shishir2009}, but no comprehensive analysis of experiments has been reported, which is the topic of this Letter.

Graphene transistors operating in the small source-drain bias V$_{\rm sd}$ regime are described by the Drude model, where the current density $j=neu=ne\mu E$, depends on the electron density n, the transverse electric field $E=V_{\rm sd}/L$ and the mobility $\mu$. At large electron drift velocity, inelastic collisions with phonons become increasingly frequent, eventually leading to a saturated $u_{\rm sat}$. It is in this velocity (current) saturated regime where analog amplifiers operate. Recent high-field transport measurements have demonstrated a high current density j of a few mA/$\mu$m in graphene transistors, with a few V on $V_{\rm sd}$. This remarkable current-carrying capability is comparable to that of carbon nanotubes (CNTs)~\cite{Yao2000, Javey2003, Park2004} and exceeds the performance of silicon transistors~\cite{SZE2006}. In contrast to CNTs, where the high-bias current in sufficiently long tubes reaches a full saturation of approximately 25~$\mu$A~\cite{Yao2000,Javey2003,Park2004}, the current in graphene does not yet fully saturate~\cite{Barreiro2009, Freitag2009}, except in the presence of a carrier density gradient~\cite{Meric2008}. The saturation in CNTs can be described by a simple phenomenological model considering instantaneous emission of CNT optical phonons of $\hbar\omega\approx 200$ meV and zone-boundary phonons of $\hbar\omega\approx 160$ meV~\cite{Yao2000, Park2004}, although alternative interpretations involving the SO phonons of the SiO$_2$ substrate are also plausible~\cite{Rotkin2009}. 
For graphene, several phonon scattering mechanisms, including the acoustic phonons of graphene~\cite{Chen2008b,Hwang2008} and the SO phonons~\cite{Wang1972} of the SiO$_2$ substrate~\cite{Fratini2008,Chen2008b} have been shown to affect low-bias transport. In the high-bias regime, both optical phonons of graphene~\cite{Barreiro2009, Bistritzer2009a} and the SO phonons of the SiO$_2$ substrate~\cite{Meric2008,Freitag2009} have been individually proposed and used to analyze experiments, although a comprehensive study combining all scattering channels and a careful evaluation of their individual contributions is still lacking. The closely related issue of dissipation of hot carrier energy becomes critical for devices operating in the high-curent regime. Recent modeling points to the crucial role of the SiO$_2$ substrate in graphene and CNT devices, although the details of the dissipation mechanism are yet to be articulated~\cite{Freitag2009,Shi2009}.

\begin{figure}
\includegraphics[width=\columnwidth]{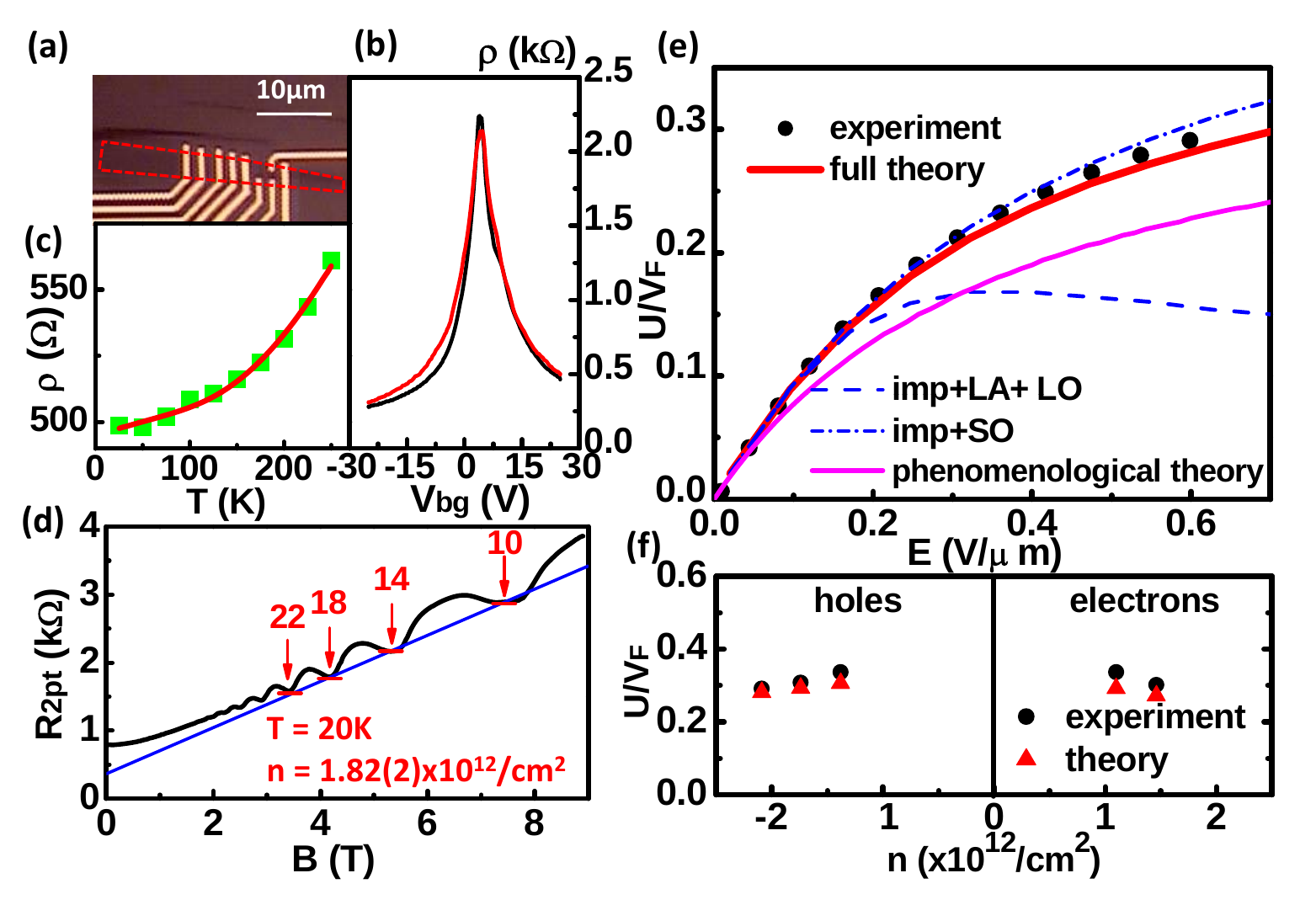}
%\vspace{-0.1in}
\caption{(Color Online) Field effect transistor device configurations and transport properties. (a) Optical micrograph of a device on SiO$_{2}$/doped Si substrate. The graphene flake is outlined in red. (b) Resistivity $\rho(V_{\rm g})$ of a device before (black) and after (red) high-bias measurements. (c) Temperature-dependent resistivity $\rho(T)$ at $n = 1.30(2)\times 10^{12}$~cm$^{-2}$. The coupling strength of LA (graphene) and SO (SiO$_{2}$ substrate) phonons are extracted from the fitting (solid line). See Ref.~\onlinecite{Zou2009} for details. (d) Two-terminal magnetoresistance $R_{\rm 2pt}$ vs magnetic field. The contact resistance is $R_{\rm con}=340\Omega$, determined by extrapolating the magnetoresistance at the quantum hall plateaus, $R_{\rm 2pt}(B)=R_{\rm con}+\frac{h}{e^2\nu}$ (with the integer filling factor $\nu$ shown on the figure), to $B=0$. $n = 1.82(2)\times 10^{12}$~cm$^{-2}$. $T=20$~K unless otherwise noted. (e) The measured (circles) and calculated (lines) drift velocity (in units of Fermi velocity, $v_{\rm F}$) vs. electric field. The conduction  of our devices tends to drop sharply and irreversibly near $E\approx 1$~V/$\mu$m, presumably due to burning at local hot spots. Dashed line: theory with only impurities and the LA and LO phonons of graphene. Dash-dotted line: theory with only impurities and the SO phonons of the SiO$_{2}$ substrate. Solid line: the full theory including all scattering mechanisms. The density of holes is $n=2.09(2)\times 10^{12}$~cm$^{-2}$, the density of charged impurities is $n_{\rm imp}=5.8\times 10^{11}$~cm$^{-2}$.
Experimental error bar is smaller than the size of the symbol. We fix the substrate temperature at $T_{\rm s}$=20 K and vary the graphene lattice temerature $T_{\rm L}$ with $E$ from 20-380~K in this calculation (with $T_{\rm L}$ as high as 470~K for other measurements).
(f) Comparison of experimental and theoretical drift velocity for several electron (right) and hole (left) densities at $E=0.6$~V/$\mu$m.}
\label{experiment}
\end{figure}

We obtain single-layer graphene sheet through mechanical exfoliation and fabricate field effect transistors (FETs) on SiO$_{2}$/doped Si substrates using standard lithographic techniques. Two-terminal and Hall bar configurations patterned onto a single graphene sheet allow us to correlate measurements of $I_{\rm sd}(V_{\rm sd})$(I-V), the small-bias resistivity $\rho (V_{\rm bg})$ and its temperature dependence $\rho(T)$ on the same sample (Fig.~\ref{experiment}~(a)). Measurements are carried out in a He$^4$ cryostat at T=20 K unless otherwise noted. Fabrication and measurement details can be found in Refs.~\onlinecite{Hong2009} and~\onlinecite{Zou2009}. Samples used in this study show a low-density field effect mobility $\mu_{\rm FE}=(d\sigma/dn)(1/e)$ of $5,000-9,000$~cm$^{2}$/Vs, comparable to samples used in other experiments~\cite{Barreiro2009, Meric2008, Freitag2009}.

A typical I-V trace is shown in Fig.~\ref{experiment}~(e). Data used here are obtained from samples that showed negligible change in $\rho(V_{\rm bg})$ before and after I-V measurements (Fig.~\ref{experiment}~(b)). I-V data are indepdent of the sweeping rate of $V_{\rm sd}$, indicating that the sample reached an equilibrated state during the sweep.
The two-terminal configuration shown in Fig.~\ref{experiment}~(a) faciliates a uniform current flow between the source and drain contacts and hence an accurate determination of $j$. It is crucial, however, to account for the contact resistance, $R_{\rm con}$, in deducing the electric field $E=(V_{\rm sd}-IR_{\rm con})/L$ and the Joule heat $P=I\times (V_{\rm sd}-IR_{\rm con})$. $R_{\rm con}$ is determined from magnetotransport data, as shown in Fig.~\ref{experiment}(d), and lies in the range of 250-350~$\Omega$, where approximately 150~$\Omega$ is due to the cryostat wiring and several tens of $\Omega$s may come from the patterned Au electrodes. Given its smallness and ohmic nature at low biases, it is a reasonable assumption that the 
Au-graphene interface resistance remains constant during our I-V measurements. The drift velocity, $u$, is obtained directly from the current, $I$, using $u=j/ne$ and $j=I/W$ where $W$ is the width of our samples.

%\vspace{-0.1in}

Our theoretical treatment of high field transport in graphene is based on the
Boltzmann equation,
\begin{equation}\label{boltzmann}
-\frac{e\vec{E}}{\hbar}\cdot\nabla_{k}f_{\bf{k}\alpha}=S_{\rm col}=S^{\rm LA}_{\rm col}+S^{\rm LO}_{\rm col}+S^{\rm SO}_{\rm col}+S^{\rm imp}_{\rm col}+S^{\rm imp'}_{\rm col},
\nonumber
\end{equation}
where the distribution function is assumed to be time independent and spatially uniform. The collision term on the right originates from impurities and phonons; We model the impurity contribution with charged and neutral components, labeled as imp and imp' respectively~\cite{CastroNeto2009, Chen2008a}.
Phonons considered here are longitudinal acoustic (LA) and longitudinal optical (LO) phonons of graphene, and the surface optical (SO) phonons of the substrate~\cite{Wang1972,Fratini2008,Chen2008b}.  The electron-electron interaction is included implicitly by choosing the ``displaced'' distribution function, explained below. The collision integral for phonons, $S_{\rm col}$, has the form
\begin{equation*}
S_{\rm col}=-\sum_{{\bf p}\gamma}\left[f_{{\bf k}\alpha}(1-f_{{\bf p}\gamma})W_{{\bf kp}}^{\alpha\gamma}-f_{{\bf p}\gamma}(1-f_{{\bf k}\alpha})W_{\bf{pk}}^{\gamma\alpha}\right],\label{Scol}
\end{equation*}
where $W_{\bf{kp}}^{\alpha\gamma}=(2\pi/\hbar)\sum_{{\bf q}s}\delta_{{\bf q}+{\bf k}-{\bf p}} \lvert M_{{\bf kp}}^{\alpha\gamma}\rvert ^{2}(N_{q}+\frac{1}{2}-\frac{s}{2})\delta(\varepsilon_{k\alpha}-\varepsilon_{p\gamma}+s\hbar\omega_{q})$, with $s=+ 1$ ($s=-1$) is for phonon absorption (emission) and $M_{{\bf kp}}^{\alpha\gamma}$ is the matrix element for the scattering process which takes an electron of momentum ${\bf k}$ in band $\alpha$ to momentum ${\bf p}$ in band $\gamma$. The quantities $N_{q}$ and $\hbar\omega_{q}$ denote the phonon occupation factor and the phonon energy. We assume the equilibrium Bose-Einstein distribution for phonons, allowing for elevated temperatures for the graphene lattice and the substrate, but ignoring non-equilibrium hot phonon effects~\cite{hotphonon}. The net charge density relative to the neutral Dirac point is given by
\begin{equation*}\label{density}
n=n_e-n_h=g\sum_{{\bf k}}\left[f_{{\bf k}c}-(1-f_{{\bf k}v})\right]
\end{equation*}
where $g=4$ arises from the spin / valley degeneracy. 

It is estimated that for large electron densities, the electron-electron scattering time is sufficiently short~\cite{Sun2008} that electrons come to equilibrium before any other scattering processes occur. The rapid establishment of an equilibrium electron distribution is approximated by assuming a displaced Fermi-Dirac distribution for the electrons,
\begin{equation*}\label{fk}
f_{{\bf k}\alpha}=[\exp(\varepsilon_{k\alpha}-\hbar{\bf u}\cdot{\bf k}-\mu_{\rm e})/k_{\rm B}T_{\rm e}+1]^{-1},
\end{equation*}
where $\alpha=\pm 1$ denotes the conduction or valance band, $\varepsilon_{k\alpha}=\alpha v_{\rm F}k$ is the energy spectrum of graphene ($v_{\rm F}=10^{8}$~cm/s), 
$\mu_{\rm e}$ is the chemical potential, $T_{\rm e}$ is the temperature of the electrons in the moving frame of reference, and ${\bf u}$ is the drift velocity. This approximation is justified a posteriori for the carrier densities reported here.  
Following Bistritzer and MacDonald~\cite{Bistritzer2009a}, we solve the equations for momentum loss rate, $Q=-en{\bf E}=-g\sum_{{\bf k}\alpha}{\bf k}S_{\rm col}$, and power dissipation, $P=-en{\bf E}\cdot{\bf u}=-g\sum_{{\bf k}\alpha}\varepsilon_{k\alpha}S_{\rm col}$, (which can be derived from the Boltzmann equation) for the three variables ${\bf u},\, \mu_{\rm e},\, T_{\rm e}$ to obtain various transport coefficients.

The concentration of charged and neutral impurities, $n_{\rm imp}$ and $n_{\rm imp'}$, are extracted from a global fit of the density dependence of the low-temperature resistivity $\rho (V_{\rm bg})$ to the charged impurity model~\cite{Chen2008a,impurities}. Small discrepancies between the global fit and the actual data are corrected for by adjusting $n_{\rm imp}$ slightly for each carrier density, resulting in 
$n_{\rm imp}\sim 4.4$-$4.6 \times 10^{11}$~cm$^{-2}$ for holes and $n_{\rm imp}\sim 5.3$-$5.8 \times 10^{11}$~cm$^{-2}$ for electrons, the difference between the two attributable to the asymmetry of the conductivity. A few percent variation in $n_{\rm imp}$ for electrons or for holes has negligible effect on the saturation velocity.  The contributions of the neutral impurities to the residual resistivity are $\rho^{\rm imp'}=12.9$~$\Omega$ for holes and $9.7$~$\Omega$ for electrons.

\begin{figure}
\includegraphics[width=\columnwidth]{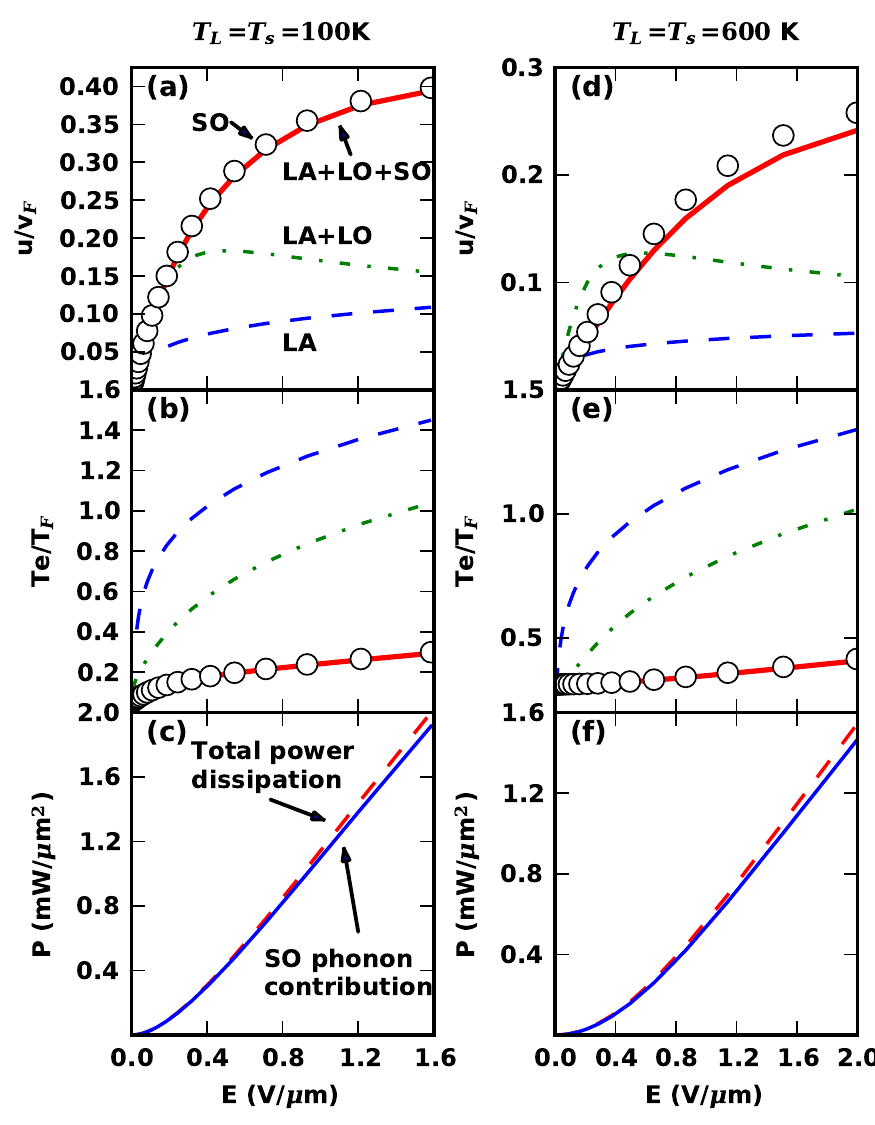}
\caption{(Color Online) The electron drift velocity, electron temperature and power dissipation in graphene field effect transistors. Panels (a), (b), and (c) consider a low-temperature device, with $T_{\rm L}=T_{\rm s}=100$~K, whereas the remaining panels assume $T_{\rm L}=T_{\rm s}=600$~K, which corresponds to a typical room-temperature device with current induced heating. In panels (a), (b), (d), and (e), results are shown including all phonons (solid lines); LA and LO phonons (dash-dotted lines); LA phonons (dashed lines); and SO phonons (open circles). Panels (c) and (f) show total power dissipation (dashed line) and also the power dissipation into the SO phonons (solid line).
For all plots, we take carrier density $n=2\times 10^{12}$ cm$^{-2}$, density of charged impurities $n_{\rm imp}=5\times 10^{11}$ cm$^{-2}$, and neglect neutral impurities.}
\label{mechanisms}
\end{figure}

The temperature dependence of the low-bias resistivity in fig.~\ref{experiment}(c) is fit to a sum of three terms. The residual resistivity is due to impurities. The linear term is attributed to acoustic phonons~\cite{Hwang2008}; we find a deformation potential $D=18$~eV, in agreement with other work~\cite{Chen2008b}. We fit the non-linear T dependence of the resistivity to the Bose-Einstein distribution for the SO phonons~\cite{Fratini2008,Chen2008b,Zou2009}. Parameters used to calculate the SO phonon frequencies are obtained from our measurements~\cite{Zou2009} and Ref.~\onlinecite{Fischetti2001}. Using the linearized Boltzmann equation with the relaxation time ansatz, and making the approximation that the electron-SO phonon scattering strength is momentum independent, we find $\rho^{SO}=\sum_{i}(\hbar/4e^{2})g_{i}^{2}(1+\hbar\omega_{i}/\varepsilon_{F})N(\omega_{i})$ in the low temperature, zero-bias limit, which we fit to the nonlinear part of the low-bias data to extract the coupling constants $g_i$ ($i=1,2$ refers to the two SO phonon modes). The low-bias data show negligible dependence on the LO phonons, so we are unable to deduce this coupling parameter from experiment, and we use instead the theoretical coupling constants from Refs.~\onlinecite{Ando2006} and~\onlinecite{Bistritzer2009}. As shown in Fig.~\ref{mechanisms}, the dominance of SO phonons renders our results insensitive to the coupling strength to the graphene LO phonons.

The results obtained from a numerical solution of the above equations are plotted in Fig.~\ref{mechanisms} for two representative sets of lattice and substrate temperatures ($T_{\rm L}$ and $T_{\rm s}$, respectively). Here we have taken $T_{\rm L}=T_{\rm s}$ for simplicity to illustrate the key features of our theory. A more sophisticated treatment of $T_{\rm L}$ is implemented in Fig.~\ref{experiment} to compare to data and the effect of $T_{\rm s}$ is examined in Ref.~\onlinecite{DaSilva_unpublished}. The most striking result is that SO phonons are the principal scattering mechanism for high field transport: the drift velocity and the electron temperature obtained from the theory including all phonons are extremely well approximated in a model that retains only the SO phonons (and impurities), and more than $\sim$95\% of the power dissipation also occurs directly into the SO phonons. 
A surprising feature is that the inclusion of SO phonons leads to an {\it increase} of the electron drift velocity, counter to the intuition that additional scattering mechanism should decrease it. The origin of this behavior lies in the fact that SO phonons also provide an efficient route for energy dissipation, leading to a drastic drop of electron temperatures, which translates into higher saturation velocities.

A detailed comparison between theory and experiment requires us to carefully consider the dependence of $T_{\rm L}$ on the applied field, $E$.  
We use an empirical relation of lattice temperature vs power dissipation, obtained from Raman spectroscopy recently~\cite{Freitag2009} to estimate $T_{\rm L}$ as a function of $E$ in our devices. $T_{\rm L}$ ranges~\cite{DaSilva_unpublished} from 20 to 380~K for the device shown in Fig.~\ref{experiment}~(e). The accuracy of $T_{\rm L}$, as well the coupling strength of the LO phonons in graphene, plays a minor role in our calculations due to the dominance of the SO phonons.  In Fig. 1, we keep the substrate temperature at the bath temperature $T_{\rm s}=T_{\rm bath}=20$~K as our calculation of the heat flow through the SiO$_{2}$ substrate indicates\cite{DaSilva_unpublished} an upper bound of $T_{\rm s}=250$~K, which produces only a small deviation ($<$~4\%) from results obtained with $T_{\rm s}=20$~K.  
There are no adjustable parameters in our calculations.

Fig.~\ref{experiment}~(e) shows that our theoretical results including either all phonons or only SO phonons agree with the measured $u(E)$ for a sample of density $n=2.09\times 10^{12}$ cm$^{-2}$ to better than 4\%. The agreement is somewhat worse for samples of lower densities, with $\sim$15\% disagreement at the smallest density shown in Fig.~\ref{experiment}~(f).  We attribute the worse agreement for small densities to complications arising from the formation of electron-hole puddles in real samples. 
We stress that a neglect of SO phonons results in substantial qualitative disagreement between theory and experiment.  A simple estimation of $u$ can be obtained by assuming instantaneous emission of the relevant optical phonon,~\cite{Mahan1987} which has been demonstrated to be quite reasonable for high field transport in CNTs~\cite{Yao2000}.
Such a treatment for graphene~\cite{Meric2008} leads to $u_{\rm sat}/v_{\rm F}=\hbar\omega_{\rm SO}/E_{\rm F}$, and results in a calculated $I=V/(R_{\rm imp}+V/I_{\rm sat})$ (dotted line), which is $\sim$20\% lower than the experimental data at $E=0.6$~V/$\mu$m.

\begin{figure}
\includegraphics[width=3.0in]{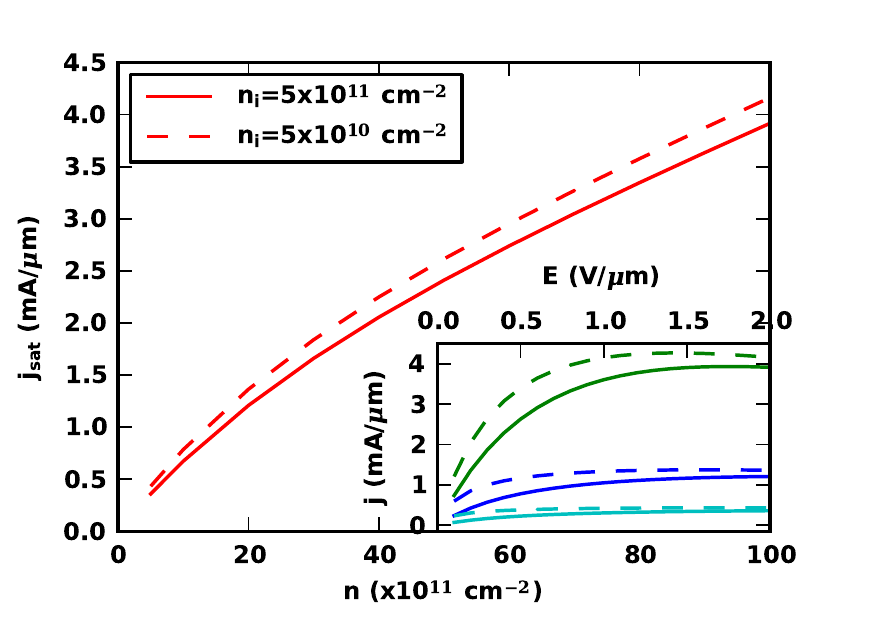}
\caption{(Color Online) Saturation current density $j_{\rm sat}$ of graphene on SiO$_{2}$. The saturation current density $j_{\rm sat}$ vs. carrier density n for charged impurity densities, $n_{\rm imp}=5\times 10^{11}$~cm$^{-2}$ (solid) and $n_{\rm imp}=5\times 10^{10}$~cm$^{-2}$ (dashed). Neutral impurities are neglected. $j_{\rm sat}$ is approximately linear for $n\gtrsim 5\times 10^{12}$~cm$^{-2}$ and is slightly lower for the higher $n_{\rm imp}$.  $j_{\rm sat}$ is defined as $j$ at $E=2$~V/$\mu$m; as seen in the inset, all currents have reached saturation at this field. Inset: current vs. electric field $j(E)$ for three representative carrier densities, $n=5\times 10^{11}$ (cyan), $2\times 10^{12}$ (blue), and $1\times 10^{13}$~cm$^{-2}$ (green). Solid and dashed curves correspond to the higher and lower $n_{\rm imp}$ respectively. All data in this figure are calculated at the representative substrate and lattice temperatures $T_{\rm s}=20$~K and $T_{\rm L}=500$~K.}
\label{prediction}
\end{figure}
%\vspace{-0.1in}
Figure~\ref{prediction} shows a theoretical prediction of saturated current density vs. carrier density in graphene transistors fabricated on SiO$_{2}$ for two charged impurity densities. Both curves display a linear regime for $n > 5\times 10^{12}$~cm$^{-2}$, where $j$ reaches a few mA/$\mu$m. These predictions point to the prospect of high-performance graphene linear amplifiers. As the sample quality improves, the current saturation will occur at lower electric field, allowing for experimental access and a greater operational range for these devices (inset). 
  
Finally, given the crucial role played by the SO phonons, it is natural to wonder if the saturation current may be enhanced by using another substrate~\cite{Hong2009,Ponomarenko2009} or a double-oxide graphene transistor.  We have studied theoretically HfO$_2$, Al$_2$O$_3$ and ZrO$_2$ substrates in vacuum/graphene/oxide structures, and find that, in spite of the large range of $\omega_{\rm SO}$ involved, the variation of the saturation velocity is less than 25\%.  We have also studied in detail, both experimentally and theoretically, the double-oxide  HfO$_2$/graphene/SiO$_2$ structure~\cite{Zou2009,DaSilva_unpublished}. The measured drift velocity is in excellent (4\%) agreement with theory, and is again dominated by the SO phonons, but, surprisingly, is {\em lower} than the drift velocity for SO$_2$ substrate alone~\cite{DaSilva_unpublished}.

In summary, by combining careful experimental and theoretical studies, we demonstrate that at high electric field, hot electrons in graphene lose energy predominately by emitting surface optical phonons of the substrate. The resulting current saturation can be accurately explained by a Boltzmann theory using experimentally obtained inputs. Cleaner samples are necessary to achieve full velocity saturation at accessible bias field strengths of $\approx 1$~V/$\mu$m. 

This work is supported by NSF CAREER DMR-0748604, NSF NIRT ECS-0609243 and the Penn State MRSEC under NSF grant DMR-0820404. The authors acknowledge the PSU site of NSF NNIN. After the submission of the manuscript, we became aware of Ref.~\onlinecite{Perebeinos2009b} which also deals with high field transport in graphene.

\end{document}